\documentclass{article}
\usepackage{spconf,amsmath,graphicx,hyperref}
\usepackage{amsmath,amssymb}
\usepackage{afterpage}
\usepackage{booktabs}  
\usepackage{array}      
\usepackage{caption}    
\usepackage{graphicx}
\usepackage{amssymb}  
\usepackage{multirow} 
\usepackage{verbatim} 
\usepackage[numbers,sort&compress]{natbib} 

\usepackage{enumitem}
\usepackage{subcaption}

\title{EMORL-TTS: Reinforcement Learning \\ for Fine-Grained Emotion Control in LLM-based TTS}
\name{%
\begin{tabular}{@{}c@{}}
\itshape
Haoxun Li$^{1}$ \quad Yu Liu$^{1}$ \quad Yuqing Sun$^{1}$ \quad Hanlei Shi$^{1}$ \quad
Leyuan Qu$^{1}$ \quad
Taihao Li$^{\star,1}$
\end{tabular}%
\thanks{$^{\star}$ Corresponding authors.}
\thanks{Haoxun Li, et al. Copyright 2026 IEEE. Personal use of this material is permitted. 
Permission from IEEE must be obtained for all other uses, including reprinting/republishing, 
creating new collective works, for resale or redistribution to servers or lists, or reuse of 
any copyrighted component of this work. DOI will be added upon IEEE Xplore publication.}
}
\address{$^{1}$ Hangzhou Institute for Advanced Study, University of Chinese Academy of Sciences}
\begin{document}
\ninept
\maketitle
\begin{abstract}
Recent LLM-based TTS systems achieve strong quality and zero-shot ability, but lack fine-grained emotional control due to their reliance on discrete speech tokens. Existing approaches either limit emotions to categorical labels or cannot generalize to LLM-based architectures. We propose EMORL-TTS (Fine-grained Emotion-controllable TTS with Reinforcement Learning), a framework that unifies global intensity control in the VAD space with local emphasis regulation. Our method combines supervised fine-tuning with reinforcement learning guided by task-specific rewards for emotion category, intensity, and emphasis. Moreover, we further investigate how emphasis placement modulates fine-grained emotion intensity. Experiments show that EMORL-TTS improves emotion accuracy, intensity differentiation, and emphasis clarity, while preserving synthesis quality comparable to strong LLM-based baselines. Synthesized samples are available on-line\footnote{\url{https://wd-233.github.io/EMORL-TTS_DEMO/}}.
\end{abstract}
\section{Introduction}
\label{sec:intro}

In recent years, Text-to-Speech (TTS) technology has advanced rapidly, with its goal extending far beyond generating ``intelligible'' speech toward achieving naturalness and expressiveness. Incorporating emotion has been shown to significantly enhance the expressive power of synthesized speech, making emotional TTS a growing research focus. Most existing studies, however, have concentrated on categorical emotion control~\cite{lee2017emotional, wu2021cross, diatlova2023emospeech, nithin2022emotional}, e.g., synthesizing speech as \textit{happy}, \textit{angry}, or \textit{sad}. Yet emotions are inherently continuous, and discrete categories fail to capture the richness of emotional strength and subtle variations~\cite{xie2025emosteer}.

To address this limitation, increasing attention has been given to \textit{emotion intensity modeling} and \textit{mixed-emotion synthesis}. For instance, Mixed Emotion~\cite{zhou2022speech} leverages relative attribute ranking to generate blended emotions; EmoMix~\cite{tang2023emomix} and EmoDiff~\cite{guo2023emodiff} employ diffusion models and soft labels to enable continuous emotion control; EmoSphere-TTS~\cite{cho2024emosphere} and EmoSphere++~\cite{cho2025emosphere++} map emotions to a three-dimensional Valence--Arousal--Dominance (VAD) sphere, where radial distance encodes intensity and angular position encodes style, providing a novel perspective for fine-grained emotion modeling. These advances highlight the importance of controllable intensity and fine-grained regulation for improving the naturalness and expressiveness of TTS.

Meanwhile, Large Language Model (LLM)-based TTS systems~\cite{wang2025spark,du2024cosyvoice,zhang2025vevo,huang2025step} have demonstrated remarkable advantages in zero-shot capability and synthesis quality, and are widely regarded as the future direction of TTS. However, most existing emotion modeling approaches are built upon non-LLM architectures, and fine-grained emotional control in LLM-based TTS remains an open challenge. A key difficulty arises because LLM-based TTS relies on discrete speech tokens rather than continuous vector representations, making it inherently difficult to directly model continuous emotion intensity or prosodic prominence. 

On the other hand, prior work such as EME-TTS~\cite{li2025eme} has demonstrated that \textit{prosodic emphasis}---the most prominent part of speech prosody---is a key factor in emotional expressiveness. Yet its method was constrained by the capacity of the model and is not applicable to the discrete token space of LLM-based architectures.

This challenge can be approached in two ways: one is to explicitly design discrete token representations that approximate fine-grained and continuous prosodic signals, which, however, typically requires extensive annotation of prosodic attributes and thus is difficult to scale; the other, as we pursue in this work, is to circumvent the limitation by employing reinforcement learning, allowing the model to implicitly discover how to regulate fine-grained emotional variation through task-specific rewards.

We unify fine-grained control into a \textbf{prosody control framework} consisting of:
\begin{itemize}[noitemsep, topsep=0pt, leftmargin=*]
    \item \textbf{Global prosody control}: modeling overall emotional intensity continuously in the VAD space;
    \item \textbf{Local prosody control}: leveraging prosodic features (pitch, energy) to determine emphasis positions, complementing and reinforcing global emotion expression.
\end{itemize}

Our method integrates Supervised Fine-Tuning (SFT) with Group Relative Policy Optimization (GRPO)~\cite{guo2025deepseek}, and introduces two task-specific rewards to guide VAD-based intensity modeling and emphasis prediction. In this way, we achieve fine-grained controllable emotion synthesis in LLM-based TTS, incorporating both global and local regulation.

The main contributions of this paper are summarized as follows:
\begin{itemize}[noitemsep, topsep=0pt, leftmargin=*]
    \item To the best of our knowledge, we are the first to introduce \textbf{VAD-based global prosody control} into LLM-based TTS, achieving continuously controllable emotional intensity with SFT and GRPO.
    \item We design a \textbf{local prosody control} mechanism based on prosodic prominence, enabling controllable emphasis positioning and enhancing fine-grained emotional regulation.
    \item We construct a \textbf{unified fine-grained emotion control framework} by combining global and local prosody control. Experiments demonstrate that our approach significantly outperforms existing methods in both synthesis quality and emotional controllability.
\end{itemize}

\begin{figure*}[!t]
\centering
\includegraphics[width=0.75\linewidth, trim=0 0 0 0, clip]{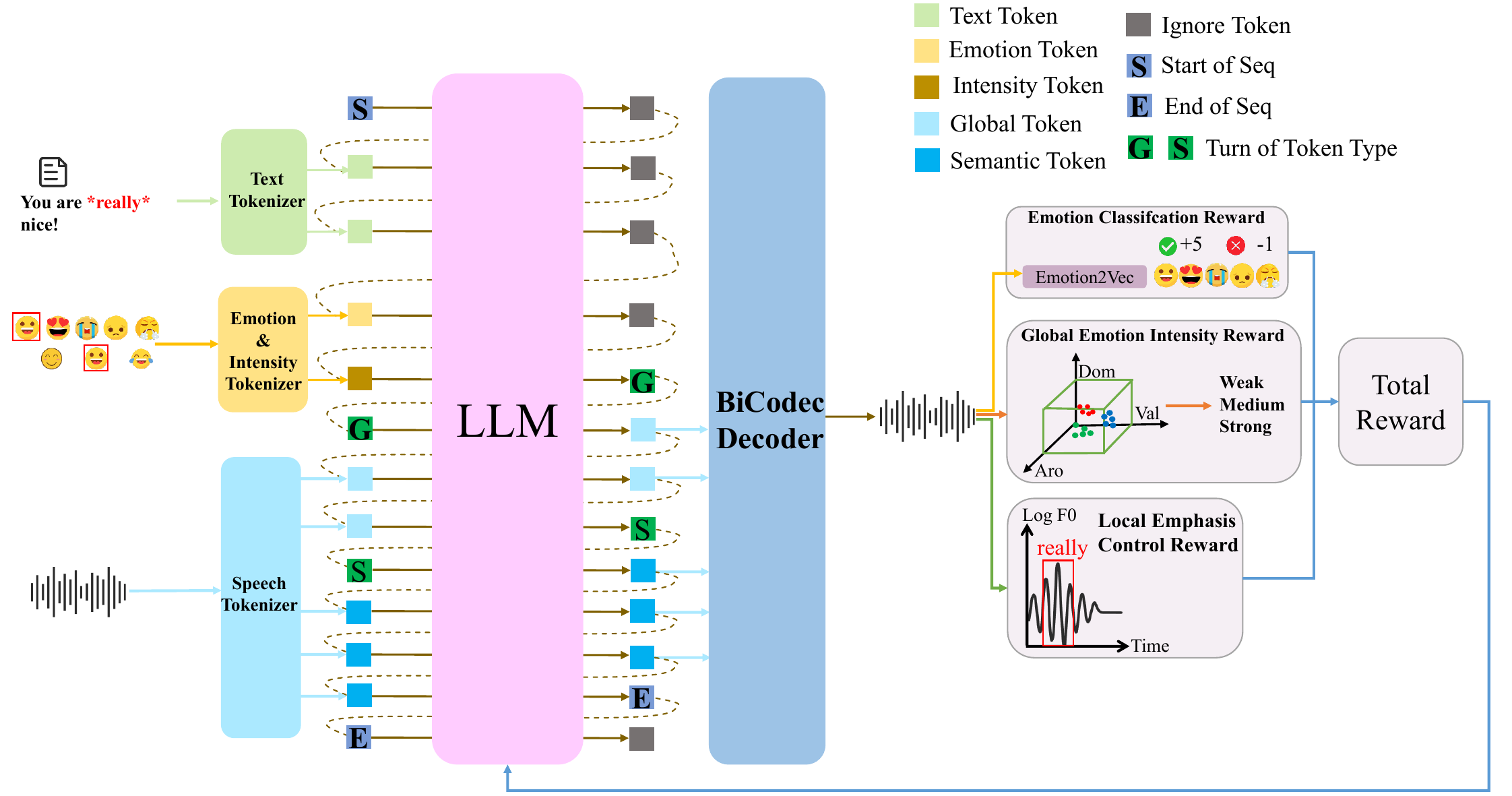}
\caption{
Overview of the proposed LLM-based fine-grained emotion-controllable TTS framework. Text, emotion, and intensity tokens are fed into the LLM, and the BiCodec decoder reconstructs the waveform. Reinforcement learning with multiple rewards (emotion classification, global emotion intensity, and local emphasis control) is employed to enhance controllability.}
\label{fig:fig_2}
\end{figure*}

\section{Method}
\label{sec:method}

\subsection{Overview}
\label{subsec:overview}

We build upon a single\mbox{-}stage LLM\mbox{-}based TTS baseline, Spark-TTS~\cite{wang2025spark}, whose \textit{BiCodec} represents speech with discrete tokens that jointly carry \emph{global acoustic traits} and \emph{semantic information}. 
Leveraging this expressive codec, we freeze the BiCodec and only adapt the LLM via a two\mbox{-}stage post\mbox{-}training paradigm: (i) SFT on emotion\mbox{-}annotated data to endow \emph{emotion\mbox{-}category controllability} while exposing the model to \emph{intensity} and \emph{emphasis} cues, and (ii) reinforcement learning with GRPO, guided by three rewards---Speech Emotion Recognition (SER) accuracy, emotion\mbox{-}intensity fidelity, and emphasis controllability---to strengthen fine\mbox{-}grained prosody control while preserving category control.

Given text input \(x\), an \emph{emotion category} \(c\in\{1,\dots,K\}\), a \emph{global intensity} cue \(r\in[0,1]\) (or discrete levels), and an optional \emph{local emphasis mask} \(m\in\{0,1\}\) that marks emphasized tokens in \(x\), the model autoregressively predicts a sequence of \emph{discrete speech tokens} \(z=(z_1,\dots,z_T)\) under a trainable LLM policy \(p_\theta\):
\begin{equation}
p_\theta\!\left(z \mid x,c,r,m\right)
=\prod_{t=1}^{T} p_\theta\!\left(z_t \mid z_{<t},\, x,\, c,\, r,\, m\right).
\label{eq:ar-factorization}
\end{equation}
A \emph{frozen} BiCodec decoder then synthesizes the waveform from tokens:
$
\hat{y} = \operatorname{BiCodecDecode}(z),
$
and only the LLM parameters \(\theta\) are updated during post\mbox{-}training.

\medskip
\subsection{Stage I: Emotion-Controllable SFT}
\label{subsec:sft}
We build upon Spark-TTS~\cite{wang2025spark}, an LLM-based TTS model with BiCodec representations, and freeze the BiCodec during post-training. The attribute tokenizer is repurposed to accept two control tokens---emotion category and discretized intensity (\textit{weak}/\textit{medium}/\textit{strong})---prepended to the text. Intensity labels are obtained from a pretrained VAD estimator~\cite{wagner2023dawn} by measuring the Euclidean distance to a neutral centroid and discretizing with category-specific thresholds; the resulting bin index is mapped to the intensity token. We fine-tune the LLM by minimizing token-level cross-entropy conditioned on these control tokens, which establishes emotion-category controllability and a calibrated intensity interface used by reinforcement learning.

\subsection{Stage II: GRPO with Multi\mbox{-}Objective Rewards}
\label{subsec:grpo}

We cast emotion\mbox{-} and emphasis\mbox{-}controllable TTS as a sequential decision process: the state \(s\in\mathcal{S}\) consists of the input text and its control tokens (emotion category and intensity), the action \(a\in\mathcal{A}\) is the generated sequence of speech tokens, and the policy \(\pi_\theta\) is the LLM of Spark\mbox{-}TTS. The training objective maximizes expected reward:
\begin{equation}
\nabla_\theta J(\theta) = \mathbb{E}_{s \sim \mathcal{D},\, a \sim \pi_\theta}\!\left[\,R(s,a)\,\nabla_\theta \log \pi_\theta(a\mid s)\right].
\label{eq:pg}
\end{equation}

\textbf{GRPO.}
For each prompt \(s\), we sample \(K\) candidates \(a^{(k)}\!\sim\!\pi_\theta(\cdot\mid s)\), compute rewards \(R^{(k)}{=}R(s,a^{(k)})\), and form a group\mbox{-}relative advantage:
$
A^{(k)} = R^{(k)} - \bar{R},\qquad \bar{R}=\tfrac{1}{K}\sum_{j=1}^{K} R^{(j)}.
$
We optimize a clipped\mbox{-}ratio objective with a KL anchor to the SFT policy \(p_{\text{SFT}}\):
\begin{equation}
\begin{aligned}
\mathcal{L}_{\text{GRPO}}(\theta)=&\mathbb{E}\!\left[\min\!\big(\rho^{(k)}A^{(k)},\,\mathrm{clip}(\rho^{(k)},1\!\pm\!\epsilon)A^{(k)}\big)\right] 
\\&\;-\;\beta\,\mathrm{KL}\!\big(\pi_\theta(\cdot\mid s)\,\Vert\, p_{\text{SFT}}(\cdot\mid s)\big),
\end{aligned}
\label{eq:grpo}
\end{equation}
where \(\rho^{(k)}=\tfrac{\pi_\theta(a^{(k)}\mid s)}{p_{\text{SFT}}(a^{(k)}\mid s)}\).

\medskip
\textbf{Rewards.}
We design three task\mbox{-}aligned terms:

\textbf{(1) Emotion Classification Reward.}
An emotion2vec\mbox{-}based SER classifier predicts \(\hat{c}=\arg\max p(c\mid \hat{y})\). To preserve category controllability acquired in SFT, we use a large, sign\mbox{-}separated shaping (and assign a higher relative weight to this term in the composite reward):
\begin{equation}
R_{\text{ser}} \;=\; 
\begin{cases}
+5, & \text{if } \hat{c}=c,\\
-1, & \text{otherwise.}
\end{cases}
\label{eq:rser}
\end{equation}

\textbf{(2) Global Emotion Intensity Reward.}
We reuse the pretrained VAD predictor~\cite{wagner2023dawn} from SFT to obtain \(\mathbf{v}(\hat{y})\in[1,7]^3\) and compute its distance to the neutral centroid
\(\boldsymbol{\mu}_{\text{neu}}=(3.8494,\,4.2614,\,3.9072)\):
\begin{align}
   d(\hat{y})=\bigl\lVert \mathbf{v}(\hat{y})-\boldsymbol{\mu}_{\text{neu}}\bigr\rVert_2. 
\end{align}
We discretize \(d(\hat{y})\) into \(\{\texttt{weak},\texttt{medium},\texttt{strong}\}\) by fixed bins, and combine a hard match with a smooth, bin\mbox{-}centered Gaussian:
\begin{equation}
\begin{aligned}
    R_{\text{match}}&=\mathbf{1}\{\mathrm{bin}(d)=r\},\\
R_{\text{dist}}&=\exp\!\left(-\frac{(d-m_r)^2}{2\sigma_r^2}\right),\\
R_{\text{int}}&=R_{\text{match}}+R_{\text{dist}},
\end{aligned}
\label{eq:rint}
\end{equation}
where \(m_r\) is the midpoint of the target bin and \(\sigma_r\) controls smoothness.

\textbf{(3) Local Emphasis Control Reward.}
We obtain word boundaries by NeMo Forced Aligner (NFA)~\cite{rastorgueva2023nemo}, and for each word \(w\in\{w_1,\dots,w_N\}\) extract prosodic features:
\begin{equation}
\begin{aligned}
f_{\text{pitch}}(w)&=\max_{\tau\in w}\log F_0(\tau)\ ,\\
f_{\text{energy}}(w)&=\text{mean}_{\tau\in w}\!\bigl\| \text{STFT}(\tau)\bigr\|_2,
\end{aligned}
\end{equation}
with a 20\,ms window. Let \(\mu_{\text{pitch}},\mu_{\text{energy}}\) be sentence\mbox{-}level means (z\mbox{-}score statistics are also computed; our soft terms use mean\mbox{-}relative deviation, equivalent to a scaled z\mbox{-}score and then clipped). For each emphasized word \(w^\star\), we define
\begin{align}
R_{\text{hard}}^{\text{pitch}} &= \mathbf{1}\{ f_{\text{pitch}}(w^\star)=\max_{w} f_{\text{pitch}}(w)\},\\
R_{\text{hard}}^{\text{energy}} &= \mathbf{1}\{ f_{\text{energy}}(w^\star)=\max_{w} f_{\text{energy}}(w)\},\\
R_{\text{soft}}^{\text{pitch}} &= \mathrm{clip}_{[-1,1]}\!\left(\frac{f_{\text{pitch}}(w^\star)-\mu_{\text{pitch}}}{\mu_{\text{pitch}}}\right),\\
R_{\text{soft}}^{\text{energy}} &= \mathrm{clip}_{[-1,1]}\!\left(\frac{f_{\text{energy}}(w^\star)-\mu_{\text{energy}}}{\mu_{\text{energy}}}\right).
\end{align}
The emphasis reward is
$
R_{\text{emp}} \;=\;
R_{\text{hard}}^{\text{pitch}} + R_{\text{hard}}^{\text{energy}} + R_{\text{soft}}^{\text{pitch}} + R_{\text{soft}}^{\text{energy}}.
$

We use the sum of the three terms:
\begin{align}
    R \;=\; R_{\text{ser}} \;+\; R_{\text{int}} \;+\; R_{\text{emp}}.
\end{align}

\section{Experiments}
\label{sec:typestyle}

\subsection{Experimental Setup}

In the SFT stage, we adopt two English emotional corpora: the Emotional Speech Database (ESD)~\cite{zhou2022emotional} and the Expresso~\cite{nguyen2023expresso} dataset. The English portion of ESD contains recordings from 10 speakers, each covering five emotions: angry, happy, sad, surprise, and neutral. Each speaker contributes 350 utterances per emotion, resulting in about 1,750 utterances and 1.2 hours of speech per speaker. From the Expresso dataset, we select the emotion-labeled subset containing 4,717 utterances annotated as happy, sad, or default, where default denotes the neutral class in the original dataset. Notably, a portion of these samples also includes emphasis annotations, which expose the model to emphasis-marked text–speech pairs during the SFT stage and provide useful prior knowledge for subsequent emphasis control. We train the model for 50 epochs with a batch size of 16 and a learning rate of 0.0002.

For the GRPO stage, we construct a text-only corpus consisting of 1,000 English sentences collected from the Internet. We randomly assign emphasis annotations to three words in each sentence to simulate diverse emphasis patterns. 
These annotated texts are then used to provide reward signals in the GRPO optimization stage, where we set the number of generations to 16, $\beta$ to 0.1, and the learning rate to $1.0 \times 10^{-6}$. All training experiments are conducted on 8 NVIDIA RTX 4090 GPUs.

\subsection{Evaluation Metrics}

The overall evaluation protocol incorporates both objective and subjective components. Objective assessments focus on speech quality and emotion accuracy, while subjective assessments are carried out through five dedicated tasks. 

A total of 30 subjects were recruited for the evaluations, and each participant was required to complete all five tasks:
\textbf{(i) Emotion Accuracy Test (EAT-EMO)}: Evaluates the correctness of emotional expression by comparing the intended target emotions with the emotions perceived by listeners;
\textbf{(ii) Emotion Intensity Test (EIT)}: Examines the ability to generate distinguishable intensity levels by asking listeners to identify the stronger sample in pairwise comparisons of weak, medium, and strong emotional speech;      
\textbf{(iii) Emphasis Accuracy Test (EAT)}: Assesses the consistency between the predicted emphasis positions and those perceived by human listeners;    
\textbf{(iv) Mean Opinion Score (MOS) Rating}: Measures the perceived naturalness and overall quality of synthesized speech on a five-point scale;
\textbf{(v) Part-of-Speech Emphasis Test (POSET)}: Investigates the effect of emphasis placement across different word categories, where participants rank the synthesized variants by perceived emotion intensity. 


To verify the emotional accuracy of EMORL, we conducted both objective and subjective evaluations. For CosyVoice2~\cite{du2024cosyvoice}, synthesis was performed with the CosyVoice2-0.5B-Instruct model using a neutral reference speaker and textual emotion prompts. For Emosphere++~\cite{cho2025emosphere++} and EMORL, all utterances were generated under medium intensity.

For objective evaluation, emotional accuracy was measured using the Emotion2vec-plus-large model \cite{ma2023emotion2vec} on 500 synthesized samples per model. For subjective evaluation, task 1 (EAT-EMO) required participants to recognize the emotions of 100 shuffled samples, with accuracy computed from binary judgments.

\begin{table}[ht]
\vspace{-4pt}
\caption{Objective Evaluation on Emotion Accuracy.  }
\label{tab:objective_evaluation}
\centering

\resizebox{0.46\textwidth}{!}{  
\fontsize{18pt}{24pt}\selectfont  
\begin{tabular}{ l c c c c c c }
\toprule
\textbf{Model} & \textbf{Mean} & \textbf{Neutral} & \textbf{Angry} & \textbf{Happy} & \textbf{Sad} & \textbf{Surprise} \\
\midrule

CosyVoice2 \cite{du2024cosyvoice}       & 0.63 & \textbf{0.99}  & 0.56 & 0.70 &  0.48&0.44 \\
EMORL-TTS w/o GRPO        & 0.81 & 0.91 & 0.78 & 0.86 & 0.75 & 0.76 \\
EmoSpeech~\cite{diatlova2023emospeech} & 0.77 & \textbf{0.99} & 0.91 & 0.72 & 0.70 & 0.52 \\
Emosphere++~\cite{cho2025emosphere++} & 0.85 & 0.97 & \textbf{0.93} & 0.78 & \textbf{0.80} & 0.77 \\

\midrule
\textbf{EMORL-TTS}           & \textbf{0.88} & \textbf{0.99} & \textbf{0.93} & \textbf{0.91} & 0.78 & \textbf{0.81} \\
\bottomrule
\end{tabular}
}
\vspace{-4pt}
\end{table}

\begin{table}[ht]
\vspace{-4pt}
\caption{Subjective Evaluation on Emotion Accuracy.  }
\label{tab:subjective_evaluation}
\centering

\resizebox{0.46\textwidth}{!}{  
\fontsize{18pt}{24pt}\selectfont  
\begin{tabular}{ l c c c c c c }
\toprule
\textbf{Model} & \textbf{Mean} & \textbf{Neutral} & \textbf{Angry} & \textbf{Happy} & \textbf{Sad} & \textbf{Surprise} \\
\midrule

CosyVoice2 \cite{du2024cosyvoice}       & 0.55 & \textbf{0.95}  & 0.23 & 0.44 &  0.48&0.65 \\
EMORL-TTS w/o GRPO       & 0.76 & 0.84 & 0.64 & 0.88 & 0.72 & 0.74 \\
EmoSpeech~\cite{diatlova2023emospeech} & 0.78 & 0.85 & 0.51 & 0.66 & 0.75 & 0.53 \\
Emosphere++~\cite{cho2025emosphere++} & 0.74 & 0.88 & 0.90 & 0.71 & 0.75 & 0.66 \\

\midrule
\textbf{EMORL-TTS}           & \textbf{0.89} & 0.91 & \textbf{0.93} & \textbf{0.95} & \textbf{0.80} & \textbf{0.87} \\
\bottomrule
\end{tabular}
}
\vspace{-4pt}
\end{table}

Tables~\ref{tab:objective_evaluation} and~\ref{tab:subjective_evaluation} present the objective and subjective evaluations of emotion accuracy across different models. Both objective and subjective evaluations demonstrate that EMORL-TTS substantially improves emotional accuracy compared with strong baselines. Moreover, they validate that the reinforcement learning adopted in the second training stage effectively enhances the controllability of emotional categories, further strengthening the alignment between intended and perceived emotions.

\textbf{Emotion Intensity. }
For EIT, all participants were asked to select the utterance with stronger emotional intensity from each sample pair. The results are summarized in Table~\ref{tab:intensity}.
As shown in the table, EMORL achieves superior performance compared to the baseline methods in almost all comparison settings. Moreover, the model maintains stable performance across all emotion categories, demonstrating its robustness in generating speech with distinguishable intensity levels.

\begin{table}[t]
\centering
\caption{Emotion intensity recognition results.}
\label{tab:intensity}
\resizebox{1\linewidth}{!}{%
\begin{tabular}{c|c|ccc}
\hline
\multirow{2}{*}{\textbf{Emotion}} & \multirow{2}{*}{\textbf{Model}} & \multicolumn{3}{c}{\textbf{Emotion Intensity Recognition}} \\
\cline{3-5}
& & Weak$<$Medium & Medium$<$Strong & Weak$<$Strong \\
\hline
\multirow[c]{3}{*}{Angry} 
& Relative Attribute~\cite{zhou2022speech} & 0.54 & 0.54 & 0.68 \\
& Emosphere++~\cite{cho2025emosphere++}    & \textbf{0.74} & 0.78 & 0.78 \\
& \textbf{EMORL-TTS}         &0.56& \textbf{0.82} & \textbf{0.82} \\
\hline
\multirow[c]{3}{*}{Happy} 
& Relative Attribute~\cite{zhou2022speech} & 0.52 & 0.63 & 0.66 \\
& Emosphere++~\cite{cho2025emosphere++}    & 0.73 & 0.66 & 0.78 \\
& \textbf{EMORL-TTS}         & \textbf{0.78} & \textbf{0.67} & \textbf{0.80} \\
\hline
\multirow[c]{3}{*}{Sad} 
& Relative Attribute~\cite{zhou2022speech} & 0.58 & 0.54 & 0.60 \\
& Emosphere++~\cite{cho2025emosphere++}    & 0.66 & 0.56 & 0.66 \\
& \textbf{EMORL-TTS}         & \textbf{0.67} & \textbf{0.82} & \textbf{0.84} \\
\hline
\multirow[c]{3}{*}{Surprise} 
& Relative Attribute~\cite{zhou2022speech} & 0.48 & 0.60 & 0.64 \\
& Emosphere++~\cite{cho2025emosphere++}    & 0.72 & 0.72 & 0.72 \\
& \textbf{EMORL-TTS}         & \textbf{0.76} & \textbf{0.80} & \textbf{0.85} \\
\hline\hline
\multirow[c]{3}{*}{Average} 
& Relative Attribute~\cite{zhou2022speech} & 0.50 & 0.52 & 0.58 \\
& Emosphere++~\cite{cho2025emosphere++}    & 0.56 & 0.47 & 0.50 \\
& \textbf{EMORL-TTS}         & \textbf{0.71} & \textbf{0.65} & \textbf{0.72} \\
\hline
\end{tabular}}
\end{table}

\textbf{Emphasis Accuracy. }
To evaluate the clarity and stability of emphasis in synthesized speech, we conducted EAT, where participants were asked to identify the emphasized words from randomly shuffled samples generated by different models. The results in Table~\ref{tab:emphasis} show that our proposed EMORL-TTS achieves higher emphasis recognition accuracy than baseline systems, indicating that the emphasized words are more reliably perceived by listeners. Furthermore, EMORL-TTS maintains stable emphasis performance across different emotions, though some categories, such as surprise, remain relatively more challenging due to their intrinsic prosodic characteristics. Overall, these findings demonstrate that our model enhances the perceptual distinctiveness of emphasis, thereby improving fine-grained prosody control in emotional speech synthesis.

\begin{table}[ht]
\caption{Emphasis Recognition Accuracy of Different Models.}
\label{tab:emphasis}
\centering
\renewcommand{\arraystretch}{1}  
\resizebox{0.48\textwidth}{!}{  
\fontsize{18pt}{24pt}\selectfont  
\begin{tabular}{ l c c c c c c }
\toprule
\textbf{Model} & \textbf{Mean} & \textbf{Neutral} & \textbf{Angry} & \textbf{Happy} & \textbf{Sad} & \textbf{Surprise} \\
\midrule
CosyVoice2~\cite{du2024cosyvoice} & 0.35 & 0.38 & 0.27 & 0.38 & 0.34 & 0.40 \\
EME-TTS~\cite{li2025eme} & 0.73 & \textbf{0.80} & 0.70 & 0.84 & \textbf{0.77} & \textbf{0.56} \\
\midrule
\textbf{EMORL-TTS} & \textbf{0.75} & \textbf{0.80} & \textbf{0.92} &\textbf{0.87} & 0.70 &0.48 \\
\bottomrule
\end{tabular}
}
\vspace{-8pt}
\end{table}

\textbf{Speech Quality and Naturalness. }
The quality and naturalness of synthesized speech were assessed through both objective and subjective measures. For objective evaluation, we employed the NISQA predictor \cite{mittag2021deep} to estimate naturalness on a five-point scale, while for subjective evaluation, participants performed task 4 (MOS Rating), providing quality ratings for 100 randomly shuffled test samples balanced across emotions.

As shown in Table~\ref{tab:mos}, EMORL-TTS achieves quality levels comparable to strong Spark-TTS~\cite{wang2025spark} and CosyVoice2~\cite{du2024cosyvoice} baselines, despite not incorporating any quality-related reward functions during reinforcement learning. This confirms that our reinforcement learning stage, designed primarily for controllability, does not compromise synthesis quality. Furthermore, benefiting from its LLM-based framework, EMORL-TTS consistently surpasses conventional systems such as Emosphere++~\cite{cho2025emosphere++}, highlighting its ability to combine fine-grained emotional control with state-of-the-art naturalness.

\begin{table}[ht]
\vspace{-2pt}
\caption{Comparison of Models for MOS and NISQA Scores. }
\label{tab:mos}
\centering
\resizebox{0.350\textwidth}{!}{  
\begin{tabular}{ l c c }
\toprule
\textbf{Model} & \textbf{MOS (\( \uparrow \))} & \textbf{NISQA (\( \uparrow \))} \\

\midrule
Spark-TTS~\cite{wang2025spark}                  & \textbf{4.96}  & \textbf{4.15} \\
EMORL-TTS w/o GRPO                 & 4.92  & 4.11  \\
Emosphere++~\cite{cho2025emosphere++}            & 4.24 & 3.78 \\

CosyVoice2 \cite{du2024cosyvoice}                 & \textbf{4.96} & 4.14 \\

\midrule
\textbf{EMORL-TTS}                 & 4.94 & 4.11 \\

\bottomrule
\end{tabular}
}
\end{table}

\textbf{Effect of Part-of-Speech Emphasis on Emotion Intensity. }Task5 (POSET) investigated how emphasis placement on different parts of speech influences perceived emotional strength. Five sentences were constructed, each containing words from five categories: adverbs, adjectives, verbs, nouns, and others. For each sentence, emphasis was assigned to one word category at a time and synthesized under four distinct emotions, producing 20 utterances per sentence. Within each sentence–emotion group, listeners were instructed to rank the five variants from 1 (weakest) to 5 (strongest) according to perceived emotional intensity. 

\begin{figure}[t]
    \centering
    \includegraphics[width=1\linewidth]{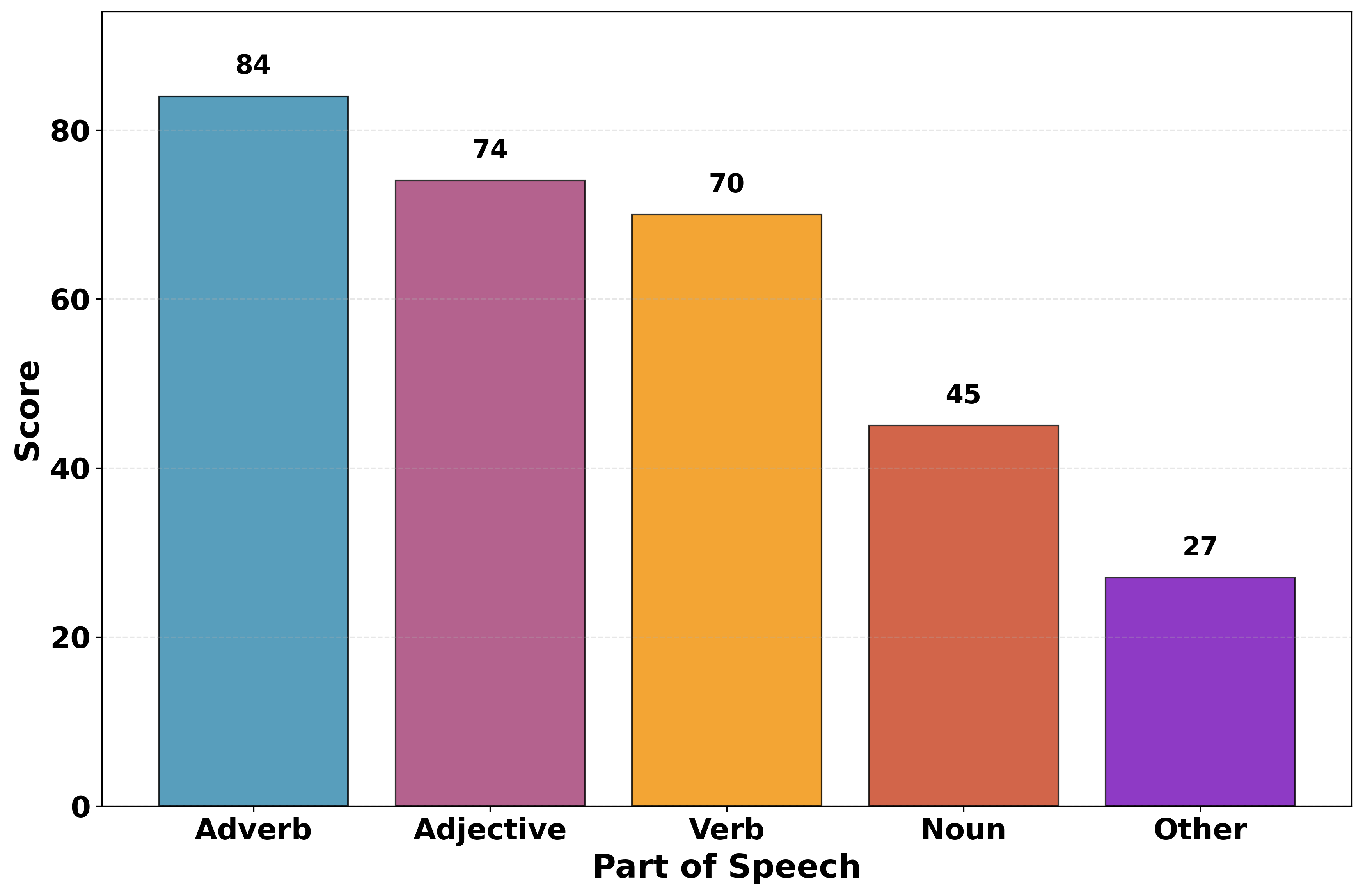}
    \caption{Aggregated emotion intensity scores across different parts of speech. Emphasis on adverbs and adjectives produces stronger perceived intensity compared to other categories.}
    \label{fig:pos_scores}
\end{figure}

Subsequently, we calculated the aggregated scores for each part of speech, as illustrated in Figure~\ref{fig:pos_scores}. The results indicate that emphasis on adverbs leads to the most pronounced enhancement of emotional intensity, followed by adjectives, while emphasis on other categories exerts relatively weaker effects. This finding suggests that strategically placing emphasis on specific word categories can serve as an effective means of achieving finer-grained control over emotional expression in synthesized speech.

\section{Conclusion}
\label{sec:typestyle}
In this work, we present a fine-grained emotion-controllable TTS framework within the LLM paradigm, tackling the challenge of modeling emotional intensity and emphasis in discrete token spaces. Combining supervised fine-tuning with reinforcement learning, and integrating global VAD-based intensity control with local prosodic emphasis, our method improves emotional accuracy, intensity differentiation, and emphasis clarity, while preserving naturalness comparable to strong LLM-based baselines. These findings show that fine-grained control is feasible in LLM-based TTS without quality loss. Future directions include cross-lingual extension, multimodal cues such as facial and gestural signals, and instruction-based controllability for more flexible expressive synthesis.

\newpage
\newpage
\section*{Acknowledgments}
This work was supported in part by the Scientific Research Starting Foundation of Hangzhou Institute for Advanced Study (2024HIASC2001), in part by the Zhejiang Provincial Natural Science Foundation of China (No.\ LQN25F020001), and in part by the Key R\&D Program of Zhejiang (2025C01104).

\noindent\textbf{Use of Generative AI and AI-Assisted Tools.}
Language editing in throughout the manuscript was assisted by ChatGPT (OpenAI) to improve grammar and clarity; all scientific content was authored by the authors.
During implementation, the authors used Cursor (an AI code assistant) for debugging support; no AI-generated code, figures, tables, or text were included in the manuscript.
All AI-assisted outputs were reviewed and verified by the authors, who take full responsibility for the content.

\section*{Compliance with Ethical Standards}
This study involved no human or animal subjects and did not require ethics approval.
\bibliographystyle{IEEEbib}
\bibliography{strings,refs}

@article{du2024cosyvoice,
  title={Cosyvoice 2: Scalable streaming speech synthesis with large language models},
  author={Du, Zhihao and Wang, Yuxuan and Chen, Qian and others},
  journal={arXiv preprint arXiv:2412.10117},
  year={2024}
}

@article{tang2023emomix,
  title={Emomix: Emotion mixing via diffusion models for emotional speech synthesis},
  author={Tang, Haobin and Zhang, Xulong and Wang, Jianzong and others},
  journal={arXiv preprint arXiv:2306.00648},
  year={2023}
}

@article{zhou2022speech,
  title={Speech synthesis with mixed emotions},
  author={Zhou, Kun and Sisman, Berrak and Rana, Rajib and others},
  journal={IEEE Transactions on Affective Computing},
  volume={14},
  number={4},
  pages={3120--3134},
  year={2022},
  publisher={IEEE}
}

@inproceedings{guo2023emodiff,
  title={Emodiff: Intensity controllable emotional text-to-speech with soft-label guidance},
  author={Guo, Yiwei and Du, Chenpeng and Chen, Xie and others},
  booktitle={ICASSP 2023},
  pages={1--5},
  year={2023},
  organization={IEEE}
}

@article{cho2024emosphere,
  title={Emosphere-tts: Emotional style and intensity modeling via spherical emotion vector for controllable emotional text-to-speech},
  author={Cho, Deok-Hyeon and Oh, Hyung-Seok and Kim, Seung-Bin and others},
  journal={arXiv preprint arXiv:2406.07803},
  year={2024}
}

@article{cho2025emosphere++,
  title={EmoSphere++: Emotion-controllable zero-shot text-to-speech via emotion-adaptive spherical vector},
  author={Cho, Deok-Hyeon and Oh, Hyung-Seok and Kim, Seung-Bin and others},
  journal={IEEE Transactions on Affective Computing},
  year={2025},
  publisher={IEEE}
}

@article{wang2025spark,
  title={Spark-tts: An efficient llm-based text-to-speech model with single-stream decoupled speech tokens},
  author={Wang, Xinsheng and Jiang, Mingqi and Ma, Ziyang and others},
  journal={arXiv preprint arXiv:2503.01710},
  year={2025}
}

@article{li2025eme,
  title={EME-TTS: Unlocking the Emphasis and Emotion Link in Speech Synthesis},
  author={Li, Haoxun and Qu, Leyuan and Hu, Jiaxi and others},
  journal={arXiv preprint arXiv:2507.12015},
  year={2025}
}

@article{diatlova2023emospeech,
  title={Emospeech: Guiding fastspeech2 towards emotional text to speech},
  author={Diatlova, Daria and Shutov, Vitaly},
  journal={arXiv preprint arXiv:2307.00024},
  year={2023}
}

@article{ma2023emotion2vec,
  title={emotion2vec: Self-supervised pre-training for speech emotion representation},
  author={Ma, Ziyang and Zheng, Zhisheng and Ye, Jiaxin and others},
  journal={arXiv preprint arXiv:2312.15185},
  year={2023}
}

@article{zhou2022emotional,
  title={Emotional voice conversion: Theory, databases and esd},
  author={Zhou, Kun and Sisman, Berrak and Liu, Rui and others},
  journal={Speech Communication},
  volume={137},
  pages={1--18},
  year={2022},
  publisher={Elsevier}
}

@article{nguyen2023expresso,
  title={Expresso: A benchmark and analysis of discrete expressive speech resynthesis},
  author={Nguyen, Tu Anh and Hsu, Wei-Ning and d'Avirro, Antony and others},
  journal={arXiv preprint arXiv:2308.05725},
  year={2023}
}

@article{mittag2021deep,
  title={Deep learning based assessment of synthetic speech naturalness},
  author={Mittag, Gabriel and M{\"o}ller, Sebastian},
  journal={arXiv preprint arXiv:2104.11673},
  year={2021}
}

@inproceedings{rastorgueva2023nemo,
  title={Nemo forced aligner and its application to word alignment for subtitle generation},
  author={Rastorgueva, Elena and Lavrukhin, Vitaly and Ginsburg, Boris},
  booktitle={Proc. Interspeech},
  year={2023}
}

@article{guo2025deepseek,
  title={Deepseek-r1: Incentivizing reasoning capability in llms via reinforcement learning},
  author={Guo, Daya and Yang, Dejian and Zhang, Haowei and others},
  journal={arXiv preprint arXiv:2501.12948},
  year={2025}
}

@article{zhang2025vevo,
  title={Vevo: Controllable zero-shot voice imitation with self-supervised disentanglement},
  author={Zhang, Xueyao and Zhang, Xiaohui and Peng, Kainan and others},
  journal={arXiv preprint arXiv:2502.07243},
  year={2025}
}

@article{xie2025emosteer,
  title={EmoSteer-TTS: Fine-Grained and Training-Free Emotion-Controllable Text-to-Speech via Activation Steering},
  author={Xie, Tianxin and Yang, Shan and Li, Chenxing and others},
  journal={arXiv preprint arXiv:2508.03543},
  year={2025}
}

@article{lee2017emotional,
  title={Emotional end-to-end neural speech synthesizer},
  author={Lee, Younggun and Rabiee, Azam and Lee, Soo-Young},
  journal={arXiv preprint arXiv:1711.05447},
  year={2017}
}

@article{wu2021cross,
  title={Cross-speaker emotion transfer based on speaker condition layer normalization and semi-supervised training in text-to-speech},
  author={Wu, Pengfei and Pan, Junjie and Xu, Chenchang and others},
  journal={arXiv preprint arXiv:2110.04153},
  year={2021}
}

@inproceedings{nithin2022emotional,
  title={Emotional speech synthesis using end-to-end neural TTS models},
  author={Nithin, SK and Prakash, Jay},
  booktitle={2022 18th International Computer Engineering Conference (ICENCO)},
  volume={1},
  pages={1--7},
  year={2022},
  organization={IEEE}
}

@article{huang2025step,
  title={Step-audio: Unified understanding and generation in intelligent speech interaction},
  author={Huang, Ailin and Wu, Boyong and Wang, Bruce and others},
  journal={arXiv preprint arXiv:2502.11946},
  year={2025}
}

@article{wagner2023dawn,
  title={Dawn of the transformer era in speech emotion recognition: closing the valence gap},
  author={Wagner, Johannes and Triantafyllopoulos, Andreas and Wierstorf, Hagen and others},
  journal={IEEE Transactions on Pattern Analysis and Machine Intelligence},
  volume={45},
  number={9},
  pages={10745--10759},
  year={2023},
  publisher={IEEE}
}

\end{document}